\documentclass[smallextended]{svjour3}      
\smartqed  

\usepackage{graphicx,graphics,times,bm,bbm,bbold,amssymb,amsmath,mathrsfs,amsfonts,dsfont,hyperref,color}
\usepackage{soul,xcolor}
\setstcolor{red}
\hypersetup{colorlinks,linkcolor=blue,citecolor=blue,urlcolor=blue}
\journalname{Quantum Information Processing}
\begin{document}
\title{Mimicking the Hadamard discrete-time quantum walk with a time-independent Hamiltonian}
\author{J. Khatibi Moqadam \and
        M. C. de Oliveira \and}
\institute{J. Khatibi Moqadam \at
              Department of Physics, Sharif University of Technology, 14588, Tehran, Iran\\
              \email{jalilkhm@physics.sharif.edu}
           \and
           M. C. de Oliveira \at
              Instituto de F\'\i sica Gleb Wataghin, Universidade Estadual de Campinas,
              13083-970, Campinas, SP, Brazil\\
              \email{marcos@ifi.unicamp.br}}
\maketitle

\begin{abstract}
The discrete-time quantum walk dynamics can be generated by a time-dependent Hamiltonian, repeatedly
switching between the coin and the shift generators. We change the model and consider the case where
the Hamiltonian is time-independent, including both the coin and the shift terms in all times.
The eigenvalues and the related Bloch vectors for the time-independent Hamiltonian are then compared
with the corresponding quantities for the effective Hamiltonian generating the quantum walk dynamics.
Restricted to the non-localized initial quantum walk states, we optimize the parameters in the
time-independent Hamiltonian such that it generates a dynamics similar to the Hadamard quantum walk.
We find that the dynamics of the walker probability distribution and the corresponding standard
deviation, the coin-walker entanglement, and the quantum-to-classical transition of the discrete-time
quantum walk model can be approximately generated by the optimized time-independent Hamiltonian. We,
further, show both dynamics are equivalent in the classical regime, as expected.
\end{abstract}

\section{Introduction}
Quantum walks, the quantum generalizations of classical random walks \cite{Portugal2013quantum} have attracted growing attentions due to their applications in quantum information processing and quantum simulation. The discrete-time quantum walk (DTQW) is a well-known model for quantizing random walks which has extensively been studied \cite{venegas2012quantum}.
Besides being useful for designing efficient quantum search algorithms \cite{Portugal2013quantum}, the model is also universal for quantum computation~\cite{lovett2010universal}. The DTQW model was used for quantum simulation of the dynamics of a charged particle in the presence of external electric fields \cite{genske2013electric,cedzich2013propagation} and the topological phases of matter \cite{kitagawa2010exploring,kitagawa2012topological,asboth2012symmetries,asboth2013bulk,%
obuse2015unveiling,cedzich2016bulk}.

{\color{black}
Quantum walks, moreover, were employed in the area of quantum foundations, in particular, in the derivation of quantum field theories, for example, the Weyl and the Dirac fields. In that context, quantum walks are considered as a special class of the more general model quantum cellular automaton \cite{schumacher2004reversible,arrighi2012partitioned}.
In fact, the large scale approximation of the automaton dynamics over Cayley graphs of Abelian groups is considered. The quantum fields emerge by imposing some general requirements such as unitarity, locality, homogeneity and isotropy on the dynamics, and in the limit of small wave vectors \cite{dariano2014derivation,Bisio2015weyl}.
The general solution for the quantum fields is obtained in the position space by using a discrete version of the Feynman's path integral \cite{feynman2010quantum}, summing over all the paths joining the desired nodes \cite{dariano2014path,dariano2015discrete,dariano2017path}.
Quantum cellular automata are also versatile tools for investigating interacting multi-particle quantum walks in which the interactions are nonlinear in terms of the field \cite{bisio2018solutions}.
}

The spin-dependent dynamics of a spin-$1/2$ particle on a one-dimensional (1D) array, can generate the 1D DTQW dynamics. At each time step, the spin of the particle is evolved to a superposition of eigenstates, by applying the coin operator
{\color{black}
$e^{-i\theta\sigma_x}$ rotating the spin state about the $x$ axis ($\sigma_x$ is the $x$-Pauli matrix and $\hbar=1$) \cite{nielsen2010quantum}.
}
The particle is then translated conditioned on the state of the spin, using the shift operator
{\color{black}
$e^{-ix_0\sigma_zp}$, where $p$ is the momentum operator, $\sigma_z$ is the $z$-Pauli matrix and $x_0$ is the distance between the array's sites \cite{Travaglione2002}.
}
By repeating the application of those operators, the particle goes to an increasing superposition of the array sites. The corresponding probability amplitudes on different sites interfere making the particle spread ballistically.
The Hamiltonian that generates the DTQW dynamics is time-dependent (piece-wise constant), since it should be switched between the coin and the shift Hamiltonians, to generate the coin and the shift operators, respectively.
{\color{black}
To be clear, suppose the spin-rotation Hamiltonian is given by $H_\mathrm{C}=\theta\sigma_x$ and the spin-dependent translation Hamiltonian is described by $H_\mathrm{S}=x_0\sigma_z p$. To realize the dynamics, during the time at which the spin is rotated, the spin degree of freedom is decoupled from the translational degree of freedom. Having rotated the isolated spin, the spin-rotation generator is switched off and the spin-dependent translation generator is turned on.
}

{\color{black}
Besides using cellular automata for deriving the Dirac equation \cite{dariano2014derivation,Bisio2015weyl,dariano2016discrete}, it was also shown that the DTQW in the continuum limit becomes equivalent to the 1D Dirac equation \cite{strauch2006relativistic,strauch2007relativistic-effect}. 
Following the latter approach, the DTQW evolution is parameterized by setting $\theta=mx_0$, $x=nx_0$ ($n\!\in\!\mathbb{Z}$ labels the array sites) and $t=lx_0$ ($l$ is the number of step), and then the distance between the array's sites is let to go to zero. At step $l$ of the walk, as $x_0\!\to\! 0$ and the other parameters remaining finite, the DTQW evolution operator $U^l = \left[ e^{-i x_0 \sigma_z p} e^{-i mx_0 \sigma_x} \right]^{t/x_0}$ becomes
$e^{-it[\sigma_z p + m \sigma_x]}$, where the Trotter's formula is used \cite{suzuki1985decomposition,suzuki1976generalized}. By letting $p=-i\partial_x$ and considering $m$ as the mass, the Dirac equation is obtained.
}

{\color{black}
The main idea of this paper is to provide foundational insights into the DTQW model by analyzing the consequences of dismissing the time-dependency in the Hamiltonian that generates the quantum walk dynamics.
We replace the piecewise-constant Hamiltonian corresponding to the DTQW dynamics with a constant Hamiltonian.
In fact, we consider the spin rotation and the spin-dependent translation generators always switched on to generate a simultaneous dynamics of both terms, namely the SCS evolution $U_\mathrm{SCS}=e^{-ix_0\sigma_zp-i\theta\sigma_x}$.
Such a dynamics is associated with the continuous evolution of a spin-particle whose spatial dynamics is coupled to its spin degree of freedom and the spin is also continuously derived. However, we impose no restriction on the wave vectors nor on the distance between the array's sites, the conditions that were used to derive the Dirac equation.
}

{\color{black}
For the given DTQW evolution $e^{-il_0\sigma_z p} e^{-i (\pi/4) \sigma_x}$, where $\theta=\pi/4$ associated with the Hadamard coin operator, we optimize the parameter ($u_1,u_2$) such that the SCS evolution $e^{-i u_1 \sigma_z p -i u_2 \sigma_x}$ generates a dynamics as close as possible to the evolution of the Hadamard walk. Note that, either of the evolutions corresponds to a single time step of the related dynamics, hence, they can be compared at any time step and there is no difference in terms of the time scale between the two dynamics.}
We optimize the parameters such that the stroboscopic evolution of the SCS evolution, at a given period $\tau$, approximately generates the Hadamard DTQW behavior. For either of the DTQW and the SCS dynamics, a walker probability distribution can be calculated, by tracing out the state of the spin. The optimized parameters are obtained by minimizing the distance between those probability distributions, which is quantified by the Hellinger distance \cite{pollard2002user}.
It should be noted that the Hellinger distance is an appropriate merit of precision for comparing two probability distributions which was widely used in quantum information theory, for instance, in studying decoherence in quantum walks \cite{alagic2005decoherence,drezgich2009complete}, quantum tomography \cite{artiles2005invitation}, characterizing the distance in quantum states and quantum channels \cite{dajka2011distance,belavkin2005operational} and characterizing quantum correlations \cite{marian2015hellinger,roga2016geometric,suciu2015gaussian,girolami2013characterizing,chang2013remedying}.
{\color{black}
Although, the focus of the paper is on the Hadamard walk, the similar approach can be used to analyze the DTQW with any coin whose angle of rotation is in the pertinent interval $\theta \in [0, \pi/2]$. Nevertheless, we also show the behavior of the Hellinger distance in terms of the coin parameter $\theta$.
}

Many experimental setups have already been proposed to implement the DTQW model and several implementations
have also been reported~\cite{Manouchehri2014}.
In particular, the DTQW can be simulated in the phase space of a harmonic oscillator which is coupled to a
two-level system (qubit). In this method, the walker is encoded in the coherent state of the resonator whose
dynamics is conditioned on the state of the qubit.
Examples of systems explored in this regard include ion
traps~\cite{Travaglione2002,Schmitz2009,Zahringer2010},
cavity quantum electrodynamics~\cite{sanders2003quantum},
ensembles of nitrogen-vacancy centers in diamond~\cite{hardal2013discrete} and optomechanical
systems~\cite{moqadam2014quantum}.
Recently, by employing a superconducting microwave resonator coupled to a transmon qubit
(a circuit QED setup), the DTQW was realized for directly measuring the topological
invariants \cite{ramasesh2017direct,flurin2017observing}.
{\color{black}
The optimized dynamics which is considered in this paper is connected with the phase space implementation of the DTQW model. Using such an approach, we express the parameters ($u_1,u_2$) in terms of some frequencies (physical quantities) in the system. We also briefly describe the implementation of the SCS evolution in a circuit QED setup.
}


The paper is organized as follows.
After a brief review of the characteristics of the DTQW operator in Sec.~\ref{sec:II}, the time-independent Hamiltonian and the corresponding evolution are explained in Sec.~\ref{sec:III}. The dynamics of the optimized SCS dynamics is explored in Sec.~\ref{sec:IV}. In Sec.~\ref{sec:V} we analyze the SCS dynamics under the effect of decoherence. A brief discussion on the physical implementation of the SCS dyanmics and a comment on the the simulation of a DTQW with a general coin together with our conclusions are presented in Sec.~\ref{sec:VI}.

\section{DTQW model}
\label{sec:II}
To construct the 1D DTQW model consider the dynamics of a spin-$1/2$ particle with the spin basis $\{|s\rangle;s=0,1\}$ which spreads according to its spin state on a 1D array with the position basis $\{|n\rangle;n\in\mathbb{Z}\}$.
Each step of the walk is realized by applying two operators. First, each component of the spin state evolves to a superposition of the spin eigenstates.
{\color{black}
That can be done by applying the coin operator ${C}(\theta) \!=\! e^{ -i\theta\sigma_x }$ which is associated with the rotation operator $R_x(2\theta)$ around the $x$ axis by the angle $2\theta$,
}
where the $x$-Pauli matrix is given by $\sigma_{x}=\left(\begin{smallmatrix}0&1 \\1 & 0 \end{smallmatrix} \right)$, hence
\begin{equation}
\label{eq:coin}
{C}(\theta) = \begin{pmatrix}\cos\theta&-i\sin\theta \\-i\sin\theta & \cos\theta \end{pmatrix}.
\end{equation}
Next, the position state of the particle is translated conditioned on the spin state, by applying the shift operator
\begin{equation}
\label{eq:shift}
{S} \!=\! \sum_n |n+1\rangle \langle n|\otimes|0\rangle \langle 0| +
                      |n-1\rangle \langle n|\otimes|1\rangle \langle 1|.
\end{equation}
The DTQW evolution operator is, therefore, given by
${U}\!=\!{S}({\mathbb{1}}\otimes{C})$, where ${\mathbb{1}}$ is the identity of the
position Hilbert space.

The shift operator defined in the basis $\{|n\rangle\!\otimes\!|s\rangle\}$ can be expressed in the
Fourier basis which is then become diagonal \cite{Portugal2013quantum}. Considering the walk on a finite
array with $d$ sites and periodic boundary conditions, namely a cycle, the position Fourier basis is given by
$\left\{|\tilde{k} \rangle \!=\! \sum_n e^{- i \tilde{k} n } \; |n\rangle / \sqrt{d}\right\}$,
where $\tilde{k}\!=\!2\pi k/d$ and $n,k\!=\!0,\ldots,d\!-\!1$.
The effect of the shift operator on the Fourier basis vectors is
\begin{equation}
\label{eq:shift_action}
{S} \;|\tilde{k}\rangle \otimes |s\rangle = e^{i\tilde{k}(-1)^s} |\tilde{k}\rangle \otimes |s\rangle,
\end{equation}
which gives the shift operator in the Fourier basis
\begin{align}
\label{eq:shift_operator}
{S} &= \sum_k \biggl(|\tilde{k}\rangle \langle\tilde{k}| \otimes |0\rangle \langle 0| e^{i\tilde{k}} +
   |\tilde{k}\rangle \langle\tilde{k}| \otimes |1\rangle \langle 1| e^{-i\tilde{k}} \biggr) \nonumber\\
            &= \sum_k |\tilde{k}\rangle \langle\tilde{k}| \otimes e^{i\tilde{k}\sigma_z},
\end{align}
where the $z$-Pauli matrix is given by $\sigma_{z}=\left(\begin{smallmatrix}1&0 \\0 & -1 \end{smallmatrix} \right)$
{\color{black}
(see Appendix \ref{app:app1}).
}

The quantum walk evolution operator in the Fourier basis is then appears as
\begin{equation}
\label{eq:CTQW_evolution}
{U} = \sum_{k} |\tilde{k}\rangle \langle \tilde{k}| \otimes
                               e^{i \tilde{k} \sigma_z} e^{-i \theta \sigma_x}.
\end{equation}
The spin part of the evolution operator in Eq. (\ref{eq:CTQW_evolution}) that acts on the spin Hilbert
space can be written as a general rotation
\begin{equation}
\label{eq:general_rotation}
e^{i \tilde{k} \sigma_z} e^{-i \theta \sigma_x}=
e^{-i \epsilon_\theta(\tilde{k}) \bm{d}_\theta(\tilde{k}).\bm{\sigma}},
\end{equation}
in which $\cos\epsilon_\theta(\tilde{k}) = \cos\theta\cos(\tilde{k})$, and
\begin{equation}
\label{Bloch_vectors}
\bm{d}_\theta(\tilde{k}) =
\biggl[\frac{\sin\theta\cos(\tilde{k})}{\sin\epsilon_\theta(\tilde{k})},
-\frac{\sin\theta\sin(\tilde{k})}{\sin\epsilon_\theta(\tilde{k})},
-\frac{\cos\theta\sin(\tilde{k})}{\sin\epsilon_\theta(\tilde{k})}\biggr],
\end{equation}
where $\bm{\sigma}=(\sigma_x,\sigma_y,\sigma_z)$ and
$\sigma_{y}=\left(\begin{smallmatrix}0&-i \\i & 0 \end{smallmatrix} \right)$ is the $y$-Pauli matrix
{\color{black}
(see Appendix \ref{app:bloch_vectors}).
}
The eigenvalues of ${U}$ are given by $e^{i\epsilon_\theta(\tilde{k})}$ and the eigenvectors
take the form $|\tilde{k}\rangle\langle \tilde{k}| \otimes |{\pm}\bm{d}_\theta(\tilde{k})\rangle$, where
$|{\pm}\bm{d}_\theta(\tilde{k})\rangle$ are the spin eigenstates in the direction of
$\bm{d}_\theta(\tilde{k})$ in the Bloch sphere.

\begin{figure}
\includegraphics[scale=.33]{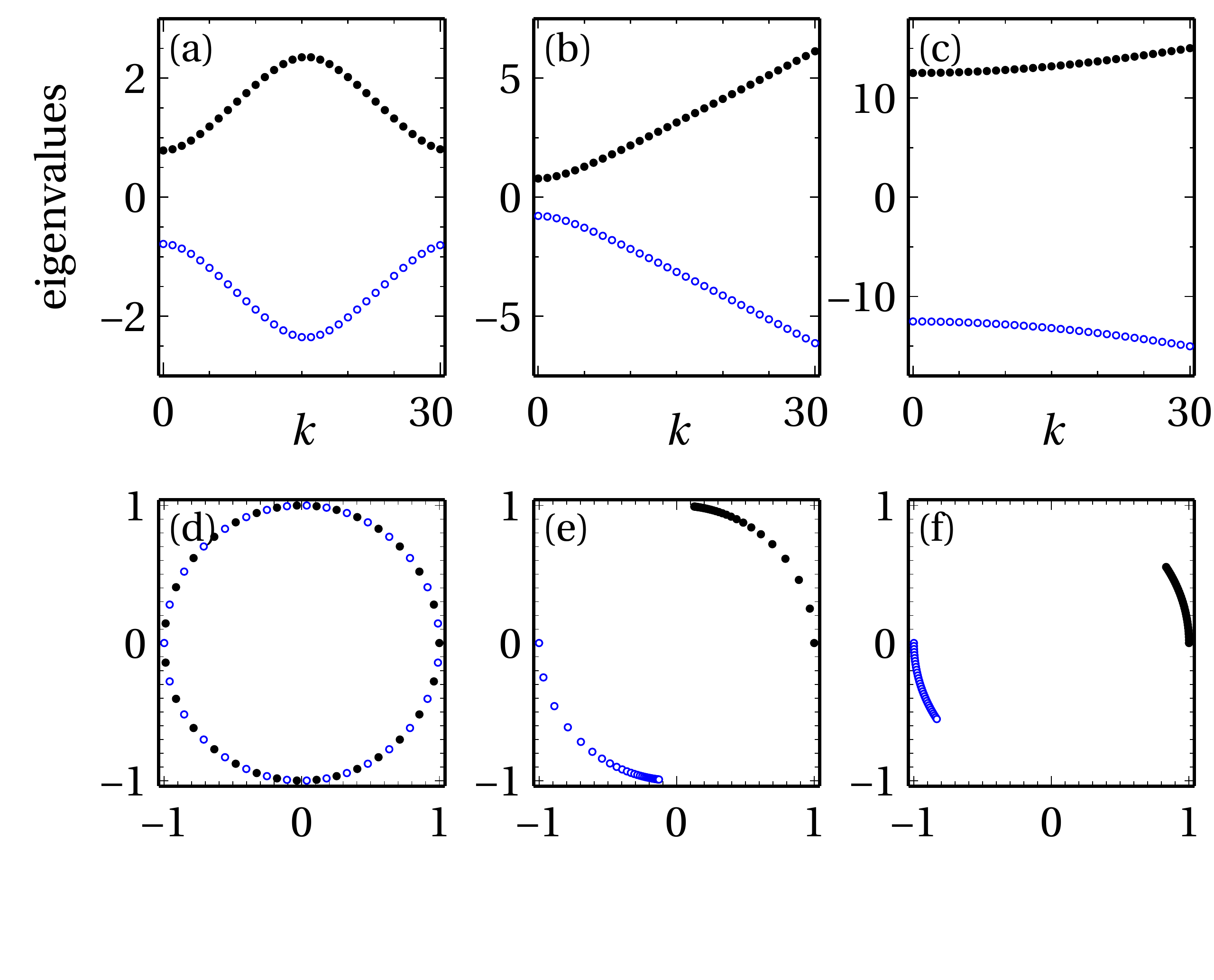}
\caption{The eigenvalues and the tip of the corresponding Bloch vectors for the DTQW operator
in (a) and (d) (first column), the SCS operator in (b) and (e) (second column) and the SCS
operator with the optimized angles in (c) and (f) (third column).
The Bloch vectors are located on a great circle of the Bloch sphere. Different colors correspond
to positive and negative eigenvalues. For the DTQW and the SCS operator $\theta\!=\!\pi/4$,
and in all cases $d=31$.}
\label{fig:eigs}
\end{figure}

The evolution operator (\ref{eq:CTQW_evolution}) can be generated by an effective Hamiltonian
in a unit time step, namely ${U}=e^{-i{H}_{\mathrm{eff}}}$, which is given by \cite{kitagawa2010exploring,kitagawa2012topological}
\begin{equation}
\label{eq:effective_H}
{H}_\mathrm{eff} = \sum_{k} |\tilde{k}\rangle \langle \tilde{k}| \otimes 
                              \epsilon_\theta(\tilde{k}) \bm{d}_\theta(\tilde{k}).\bm{\sigma},
\end{equation}
where $\tilde{k}$ and $\epsilon_\theta(\tilde{k})$ can be associated with the quasi momentums and quasi
energies of the system.
Figure \ref{fig:eigs} (a) shows the eigenvalues $\epsilon_\theta(\tilde{k})$ in terms of $k$,
for $\theta\!=\!\pi/4$ and $d\!=\!31$.
The Bloch vectors $\bm{d}_\theta(\tilde{k})$ lie on the plane perpendicular
to $(0,\cos\theta,-\sin\theta)$ and pass through the origin.
Figure \ref{fig:eigs} (b) shows the tip of the Bloch vectors on the intersection of that plane and
the Bloch sphere.
When $k$ spans its corresponding interval ($0,\ldots,d\!-\!1$), $\bm{d}_\theta(k)$
winds around the origin defining the topological invariant (winding number) of the walk, which is $1$
in this case.

Note that, the stroboscopic evolution of the quantum walk at unit time steps effectively behaves as it was
described by the time-independent Hamiltonian (\ref{eq:effective_H}). However, to physically achieve such
evolution, as described in the next section, indeed a time-dependent Hamiltonian is required.

\section{Time-independent Hamiltonian}
\label{sec:III}
Associated with the shift operator ${S}$ in Eq. (\ref{eq:shift_operator}), we can define the shift
Hamiltonian
\begin{equation}
{H}_\mathrm{S} = - \sum_{k} |\tilde{k}\rangle \langle \tilde{k}| \otimes 
                                            \tilde{k} \sigma_z,
\end{equation}
that generates the dynamics in a unit time step, ${S}\!=\!e^{-i{H}_{\mathrm{s}}}$.
In the same way, the coin Hamiltonian can be defined as
\begin{equation}
{H}_\mathrm{C} = \theta \sigma_x,
\end{equation}
that generates the coin operator in a unit time step.
Thus, if the Hamiltonian that describes the spin-$1/2$ particle on a 1D array alternates between
${\mathbb{1}} \otimes {H}_\mathrm{C}$ and ${H}_\mathrm{S}$ the DTQW dynamics is generated.
Such Hamiltonian is in fact time-dependent which is associated with the effective
Hamiltonian~(\ref{eq:effective_H}).

In this paper, however, we are interested in the dynamics of the system that is described by
\begin{equation}
\label{eq:scs_H}
{H}_\mathrm{SCS} = {H}_\mathrm{S} + {\mathbb{1}} \otimes {H}_\mathrm{C},
\end{equation}
in which both the coin and the shift Hamiltonians are applied simultaneously, hence, we call it simultaneous
coin and shift (SCS) Hamiltonian. In contrast to the DTQW, the SCS Hamiltonian (\ref{eq:scs_H}) is 
time-independent whose time evolution for a unit time step is given by
\begin{align}
\label{eq:SCS_evolution}
{U}_\mathrm{SCS} &= e^{-i{H}_\mathrm{SCS}} \nonumber\\
&=\sum_{k} |\tilde{k}\rangle \langle \tilde{k}| \otimes e^{i\tilde{k}\sigma_z-i\theta\sigma_x}.
\end{align}
As before
{\color{black}
(in a similar way as done in Appendix \ref{app:bloch_vectors}),
}
the spin part of the operator that acts on the spin Hilbert space can be written as
$$e^{i\tilde{k}\sigma_z-i\theta\sigma_x}=
e^{-i\epsilon^\mathrm{SCS}_\theta(\tilde{k}) \bm{d}^\mathrm{SCS}_\theta(\tilde{k}).\bm{\sigma}},$$
where
\begin{align}
\label{eq:scs_eigval-vec}
\epsilon^\mathrm{SCS}_\theta(\tilde{k}) &= \pm \sqrt{{\tilde{k}}^2+\theta^2}, \nonumber\\
 \bm{d}^\mathrm{SCS}_\theta(\tilde{k}) &= \biggl[ \frac{\theta}{\epsilon^\mathrm{SCS}_\theta(\tilde{k})},0,
 -\frac{\tilde{k}}{\epsilon^\mathrm{SCS}_\theta(\tilde{k})}\biggr].
\end{align}
The eigenvalues and eigenvectors of ${U}_\mathrm{SCS}$ are given by
$e^{i\epsilon^\mathrm{SCS}_\theta(\tilde{k})}$ and \linebreak
$|\tilde{k}\rangle\langle \tilde{k}| \!\otimes\!|{\pm}{\bm{d}}^\mathrm{SCS}_\theta(\tilde{k})\rangle$,
respectively, where $|{\pm}{\bm{d}}^\mathrm{SCS}_\theta(\tilde{k})\rangle \!=\!
\cos(\beta/2) |0\rangle \pm \sin(\beta/2)|1\rangle,$ in which
{\color{black}$\beta$ is the angle
}
between the positive direction of ${\bm{d}}^\mathrm{SCS}_\theta(\tilde{k})$ and the $z$ axis and
$\cos^2(\beta/2) = [1-\tilde{k}/\epsilon^\mathrm{SCS}_\theta(\tilde{k})]/2$.

Moreover, the SCS Hamiltonian in Eq. (\ref{eq:scs_H}) can be written in the form
\begin{equation}
\label{eq:scs_H_kspace}
{H}_\mathrm{SCS}= \sum_{k} |\tilde{k}\rangle \langle \tilde{k}| \otimes
               \epsilon^\mathrm{SCS}_\theta(\tilde{k}) \bm{d}^\mathrm{SCS}_\theta(\tilde{k}).\bm{\sigma}.
\end{equation}
Figure \ref{fig:eigs} (b) shows the eigenvalues $\epsilon^\mathrm{SCS}_\theta(\tilde{k})$
in terms of $k$, for $\theta\!=\!\pi/4$ and $d\!=\!31$.
The Bloch vectors $\bm{d}^\mathrm{SCS}_\theta(\tilde{k})$, for the SCS Hamiltonian, lie on the $xz$-plane
and pass through origin.
Figure \ref{fig:eigs} (e) shows the tip of the Bloch vectors on the intersection of the $xz$-plane and the
Bloch sphere. In contrast to the DTQW, when $k$ spans its interval, $\bm{d}^\mathrm{SCS}_\theta(\tilde{k})$
does not complete a round on the circle.
The winding number is $0$, hence, the SCS and the DTQW effective Hamiltonians belong to different topological
classes.

\begin{figure}
\includegraphics[scale=0.33]{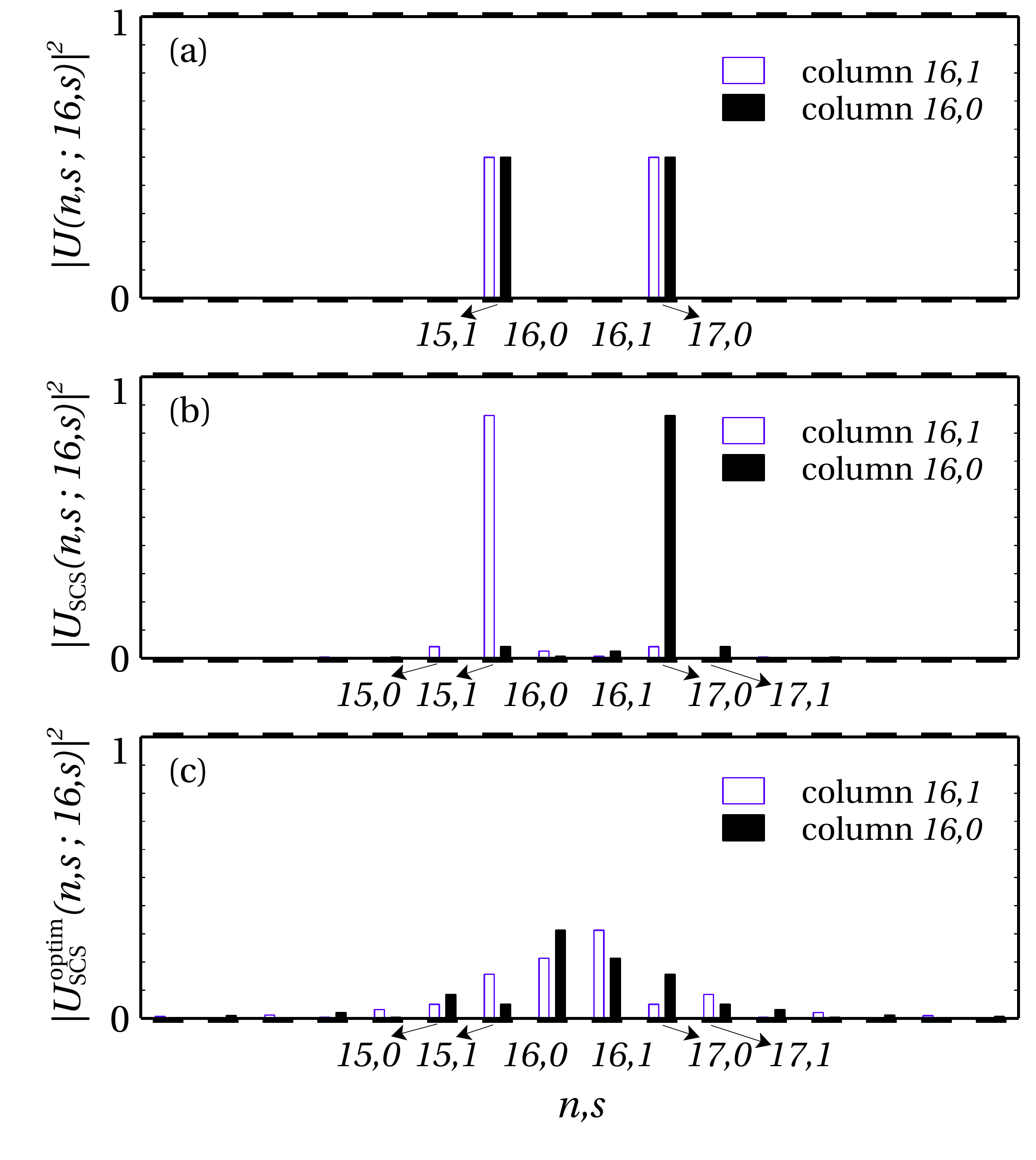}
\caption{The absolute square of the elements in the column $n\!=\!(d\!+\!1)/2,s\!=\!0,1$ of the
DTQW operator in (a), the SCS operator in (b) and the SCS operator with optimized angles in (c).
The pattern is similar for all columns $n,s$. For the DTQW and the SCS
operator $\theta\!=\!\pi/4$, and in all cases $d=31$.}
\label{fig:diag}
\end{figure}

In the position space, the DTQW evolution operator ${U}\!=\!{S}({\mathbb{1}}\otimes{C})$ transforms the basis
vectors $|n\rangle\!\otimes\!|s\rangle$ as
\begin{align}
\label{eq:dtqw_components}
{U} |n\rangle\!\otimes\!|0\rangle &\!=\!
     \cos\theta |n+1\rangle\!\otimes\!|0\rangle
   \!-\!i\sin\theta \; |n-1\rangle\!\otimes\!|1\rangle,\nonumber\\
{U} |n\rangle\!\otimes\!|1\rangle &\!=\!
     -i\sin\theta |n+1\rangle\!\otimes\!|0\rangle
    \!+\!\cos\theta |n-1\rangle\!\otimes\!|1\rangle,
\end{align}
hence, in each column of ${U}$ there are just two nonzero elements.
Figure \ref{fig:diag} (a) shows the absolute square of the elements in
columns $n\!=\!(d\!+\!1)/2,s\!=\!0,1$
of ${U}$, for $\theta\!=\!\pi/4$ and $d\!=\!31$.

{\color{black}
The effect of the SCS operator on the same basis vectors can be obtained by using Eq. (\ref{eq:SCS_evolution}) and noticing that $\langle \tilde{k} | n \rangle = e^{i\tilde{k}n}/\sqrt{d}$
}
\begin{align}
\label{eq:scs_components}
{U}_\mathrm{SCS} \; |n\rangle\!\otimes\!|s\rangle =
\sum_{n'} \bigg[ \frac{1}{d} \sum_k e^{-i\tilde{k}(n'-n)} \; |n'\rangle\!\otimes\!
e^{-i\epsilon^\mathrm{SCS}_\theta(\tilde{k})
\bm{d}^\mathrm{SCS}_\theta(\tilde{k}).\bm{\sigma}} |s\rangle \bigg],
\end{align}
which implies that the elements of the SCS operator are all nonzero, in general (note that we have used the
Fourier basis to write the expansion, but, anyway, the result is in terms of the basis
vectors $|n\rangle\!\otimes\!|s\rangle$).
Numerical simulations with $\theta\!=\!\pi/4$ show that the dominant elements of the SCS operator are
located around the diagonal of the operator.
Figure \ref{fig:diag} (b) shows the absolute square of the elements in columns $n\!=\!(d\!+\!1)/2,s\!=\!0,1$
of ${U}_\mathrm{SCS}$, for $\theta\!=\!\pi/4$ and $d\!=\!31$.
Numerical simulations (not presented here) show that changing $d$ slightly modifies the absolute square
of the elements, but, the general pattern mainly remains similar to the figure.
According to Figure \ref{fig:diag} (b),
the effect of the SCS operator, with $\theta\!=\!\pi/4$, on the basis vectors
$|n\rangle\!\otimes\!|s\rangle$, is approximately similar to the effect
of the shift operator; after the application of SCS operator on the system at the state
$|n\rangle\!\otimes\!|s\rangle$, the system goes to the state $|n+(-1)^s \rangle\!\otimes\!|s\rangle$,
with high probability.

We have defined the SCS operator in Eq. (\ref{eq:SCS_evolution}), by considering the evolution of the
time-independent Hamiltonian (\ref{eq:scs_H}) at a unite time step. In fact, the time step
can be set arbitrarily to $\tau$, hence $\tilde{k}$ and $\theta$ in Eq. (\ref{eq:SCS_evolution}) can be
replaced by $g\tau$ and $\omega\tau$, respectively, where $g$ and $\omega$ are some frequencies.
Therefore, the SCS operator can be controlled using two parameters, in contrast to the DTQW operator
which is determined by just one parameter, namely the coin rotation angle.

{\color{black}
However, before more exploring the SCS dynamics in the next section, further justifications of the SCS Hamiltonian is presented at this point. In fact, we replaced the piecewise-constant Hamiltonian that generates the DTQW dynamics with a constant Hamiltonian. The standard approach to such a situation is to use decomposition relations \cite{suzuki1985decomposition,suzuki1976generalized}. However, decomposition formulas are accurate only for sufficiently small parameters of the system, which are not corresponding to a general DTQW evolution. According to the Trotter's formula \cite{suzuki1985decomposition,suzuki1976generalized}
\begin{equation}
\label{eq:trotter}
\lim_{n\to \infty} \left[ e^{-i (\frac{1}{n}) H_\mathrm{S} }
e^{-i (\frac{1}{n}) {\mathbb{1}} \otimes {H}_\mathrm{C}} \right]^n=
e^{-i H_\mathrm{S} -i {\mathbb{1}} \otimes {H}_\mathrm{C}},
\end{equation}
which implies that quantum walks with very short distances between the array's sites and very small angles for the spin rotation, given by the left-hand side of Eq. (\ref{eq:trotter}), can be approximately generated by a constant Hamiltonian. We do not take that direction, since we demand arbitrary distances between the array's sites and large angles for the spin rotations. 
We could also use the Baker-Campbell-Hausdorff (BCH) Formula \cite{suzuki1977convergence}
\begin{equation}
\label{eq:BCH}
e^{-i \lambda H_\mathrm{S}} e^{-i \lambda {\mathbb{1}} \otimes {H}_\mathrm{C}} =
e^{-i \lambda H_\mathrm{S} -i \lambda {\mathbb{1}} \otimes {H}_\mathrm{C} \;+\; [Z(\lambda)-Z_1(\lambda)]},
\end{equation}
where $\lambda$ is a parameter, $Z(\lambda) = \sum_{n=1}^{\infty} \lambda^n Z_n$ and $Z_1 = -i \lambda H_\mathrm{S} -i \lambda {\mathbb{1}} \otimes {H}_\mathrm{C}$ (see Ref. \cite{suzuki1977convergence} for calculating the general coefficients $Z_n$). The BCH formula converges for $|\lambda|\bigl( \| H_\mathrm{S}\| + \| {\mathbb{1}} \otimes {H}_\mathrm{C} \| \bigr) < \ln(2)$, where $\|.\|$ is the standard operator norm (see Theorem 5 in Ref. \cite{suzuki1977convergence}). The convergence condition implies $|\lambda|\theta < \ln(2)$ signifying that the BCH formula does not converges for the Hadamard DTQW, for which we need $|\lambda|\theta=\pi/4$. Therefore, adding higher order terms to the exponent in the right-hand side of Eq. (\ref{eq:BCH}) does not improve its accuracy. In this way, we resort to work with the minimum Hamiltonian, namely the time-independent Hamiltonian $H_\mathrm{S}+{\mathbb{1}} \otimes {H}_\mathrm{C}$, and pick the Hadamard coin as the most interesting coin for which the BCH formula diverges.
Nevertheless, we show that even with the first order approximation, many features of the original dynamics can be recovered.
}

\section{SCS dynamics}
\label{sec:IV}
We explore the SCS dynamics referring to the realization of the DTQW in the phase space, which is
briefly explained here \cite{sanders2003quantum}.
In that realization, the walker is encoded on the coherent state of a harmonic oscillator. The shift operator
is generated by the interaction of the resonator with a qubit in the form $ga^{\dagger}a\sigma_z$, where
$a^{\dagger}$ ($a$) is the resonator creation (annihilation) operator and $g$ is the qubit-resonator
coupling strength. The coin operator can be realized by the Hamiltonian
$\omega\sigma_x$ in which $\omega$ is a frequency.
Suppose the initial state of the system is given by
\begin{equation}
\label{eq:initial_state}
|\psi_0\rangle = |\alpha\rangle|s\rangle,
\end{equation}
where
$| \alpha \rangle = e^{-|\alpha|^2/2} \sum_m\alpha^m | m \rangle / \sqrt{m!}$
is the coherent state of resonator and $|s \rangle$ is the state of qubit. The basis
$\{| m \rangle;m=0,1,\dots\}$ comprises the number states spanning the Fock space.
By setting $\omega\tau_\mathrm{C}\!=\!\theta$, for a time period $\tau_\mathrm{C}$, the coin
operator ${C}(\theta)$ is generated.
To obtain the system dynamics, the infinite dimensional Fock space is truncated and just the first $d$
eigenstates, namely the number states with fewer than $d$ photons, are kept. By
setting $g\tau_\mathrm{S}\!=\!2\pi/d$, for a time period $\tau_\mathrm{S}$,
the coherent state is rotated conditioned on the qubit state
$$e^{i(2\pi/d)a^{\dagger}a\sigma_z} |\alpha\rangle |s\rangle =
|\alpha e^{(-1)^s i (2\pi/d)}\rangle |s\rangle,$$
where $a^{\dagger}a |m\rangle \!=\! m|m\rangle$ is used. Note that the operator
$e^{i(2\pi/d)a^{\dagger}a\sigma_z}$, represented in the number state basis, is diagonal having exactly
the same form as the shift operator, given in Eq. (\ref{eq:shift_operator}).
The ``array sites" in the phase space are given by the basis
$\left\{|\varphi_n \!=\! 2\pi n/d \rangle\ \!=\! \sum_m e^{i \varphi_n m } \; |m\rangle / \sqrt{d}\right\}$
where $n,m\!=\!0,\ldots,d-1$ and the basis vectors span the truncated phase space. 

The DTQW dynamics is generated by preparing the system in the state (\ref{eq:initial_state}) and applying
the DTQW operator with a fixed coin, repeatedly. After $l$ times application, the state of the system is
obtained by using
$|\psi_l\rangle = {U}^l |\psi_0\rangle$,
and the probability distribution of the walker in the phase space basis is given by
\begin{equation}
\label{eq:probability}
P_l(\varphi_n) = \bigg| \big( \langle\varphi_n| \langle 0| \big) | \psi_l \rangle \bigg|^2
               + \bigg| \big( \langle\varphi_n| \langle 1| \big) | \psi_l \rangle \bigg|^2.
\end{equation} 

For the SCS operator
\begin{equation}
{U}_\mathrm{SCS}(\omega\tau,g\tau) = e^{ig\tau a^{\dagger}a\sigma_z -i\omega\tau\sigma_x },
\end{equation}
we fix the pair of angles $(\omega\tau,g\tau)$, and apply it $l$ times on the initial state
(\ref{eq:initial_state}). That leads to the final state
$|\psi_l^{\mathrm{SCS}}\rangle \!=\! {U}^l_\mathrm{SCS} |\psi_0\rangle$ with the phase
space probability distribution $P_l^\mathrm{SCS}(\varphi_n)$, which is obtained by replacing
${U}^l$ with ${U}^l_\mathrm{SCS}$ in Eq. (\ref{eq:probability}).

In the followings, we fix the coin to the Hadamard-like operator
\begin{align*}
{C}(\pi/4) &= e^{ -i(\pi/4)\sigma_x } \\ &= \frac{1}{\sqrt{2}} \begin{pmatrix}1&-i \\-i & 1 \end{pmatrix},
\end{align*}
for the DTQW evolution, set the initial state of the qubit in Eq. (\ref{eq:initial_state}) to the ground
state ($s=0$) and then find the pair of angles
$(\omega\tau,g\tau)$ such that the SCS operator generates similar phase probability distribution as the
Hadamard walk generates.
In fact, we minimize the difference between the probability distributions corresponding to the
DTQW and the SCS dynamics, in a given number of steps.
The difference between the two probability distributions can be measured by the
Hellinger distance, which is given by \cite{pollard2002user}
\begin{equation}
\label{eq:Hellinger}
D_\mathrm{H}(P_l,P_l^\mathrm{SCS}) = \frac{1}{\sqrt{2}}
\;\bigg\| \sqrt{P_l} - \sqrt{P_l^\mathrm{SCS}} \bigg\|_2,
\end{equation}
where $\|.\|_2$ is the Euclidean vector norm.
The angles are the solution of the optimization problem
\begin{equation}
\label{eq:optimization}
\min_{(\omega\tau,g\tau)} \frac{1}{l_0}\sum_{l=1}^{l_0} D_\mathrm{H}(P_l,P_l^\mathrm{SCS}),
\end{equation}
where the optimization is carried over $l_0$ time steps.

The optimized angles are obtained as $\omega\tau\!\approx\!(15.9462)\pi\!/\!4$ and
$g\tau \!\approx\! (1.3650)2\pi\!/\!d$, for $d\!=\!31$ and $l_0\!=\!50$, which minimize the difference
between the probability distributions for the Hadamard DTQW and the SCS dynamics.
The effects of the optimization process on the SCS spectrum and the corresponding Bloch
vectors can be viewed in Fig. \ref{fig:eigs} (c) and (f), respectively.
Comparing with the SCS operator with non-optimized parameters given in Fig. \ref{fig:eigs} (b) and (e),
the spectrum becomes more flat and transforms into two approximately straight lines with small slops,
and the distance between the positive and the negative eigenvalues increases. Moreover, the tip of
the Bloch vectors are pushed toward the points $(\pm1,0,0)$ on the Bloch sphere.
Such modifications correspond to the increase in the angle $\omega\tau=\theta$ after the optimization.
Note that, in the case $\theta\!\to\!\infty$, by using Eq. (\ref{eq:scs_eigval-vec}) we find
$\varepsilon_\theta^\mathrm{SCS}(\tilde{k})\!\to\!\pm\theta$ and
$\bm{d}^\mathrm{SCS}_\theta(\tilde{k})\!\to\!(\pm 1,0,0)$.

Figure \ref{fig:diag} (c) shows
the absolute square of the elements in columns $n\!=\!(d\!+\!1)/2,s\!=\!0,1$ of the optimized SCS
operator, denoted by ${U}_\mathrm{SCS}^\mathrm{optim}$,
for $d=31$. It can be seen that the dominant elements are distributed around the diagonal.
Having applied ${U}_\mathrm{SCS}^\mathrm{optim}$ on the system at the state
$|n\rangle|s\rangle$, the highest probable final states $|n'\rangle|s\rangle$, with $n'\!\neq\!n$, are
$|n\pm1\rangle|0\rangle$ and $|n\pm1\rangle|1\rangle$. Therefore,
${U}_\mathrm{SCS}^\mathrm{optim}$ delocalizes the initial localized state of the system,
similar to the DTQW operator as given in Eq. (\ref{eq:dtqw_components}). Of course,
when ${U}_\mathrm{SCS}^\mathrm{optim}$ is applied there is a
high probability that the system stays in its current state and also there is a low probability
that the system goes to the sites beyond the first neighbors in the position space.
Numerical simulations (not presented here) show that changing $d$ slightly modifies the absolute square
of the elements, but, the general pattern mainly remains similar to the figure.

\begin{figure}
\includegraphics[scale=.33]{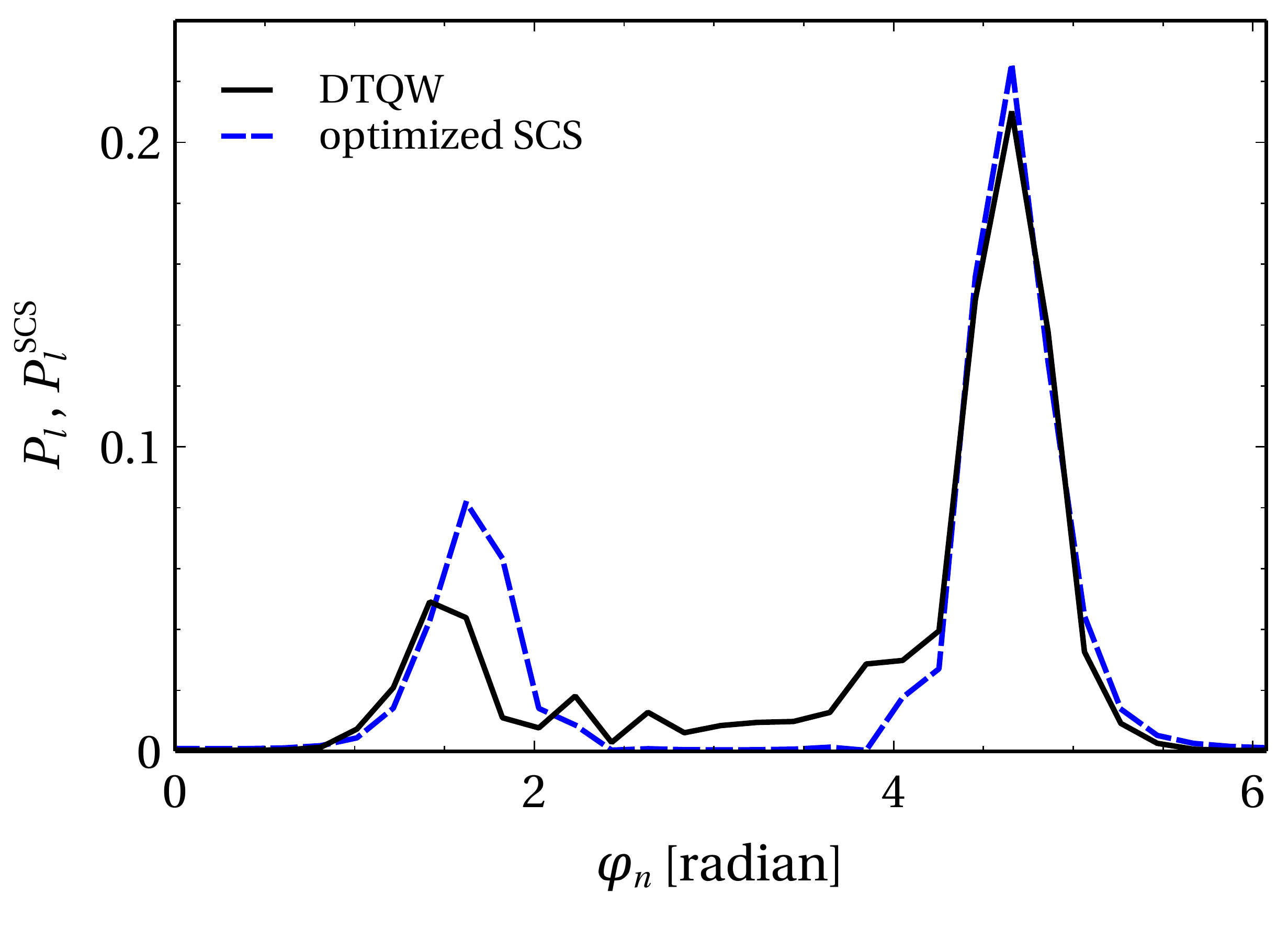}
\caption{The phase probability distributions for
the DTQW dynamics with the Hadamard-like coin (solid/black line) and the SCS dynamics with the
optimized angles (dashed/blue line).
The initial coherent state is taken as $|\alpha\!=\!5e^{i\pi}\rangle$ and the phase space dimension (the
number of sites for the walk) is set to $d\!=\! 31$. The initial state of the qubit is set to the ground
state.}
\label{fig:prob_dist}
\end{figure}

Figure~\ref{fig:prob_dist} shows the walker probability distributions at step $l\!=\!12$ for the
DTQW dynamics with the Hadamard coin, and the SCS operators with the optimized angles as described above.
Before crossing the boundaries at $\varphi\!=\!0,2\pi$, the phase probability distribution corresponding
to either of the DTQW and the optimized SCS dynamics comprises two major peaks moving in opposite directions.
Figure~\ref{fig:prob_dist} shows the situation just for one step.

Although qualitatively similar in terms of the peaks, the two plots in Fig.~\ref{fig:prob_dist} are different
regarding the intermediate interference pattern that appears for the DTQW dynamics and it is suppressed
for the optimized SCS evolution.
The smoothness in the optimized SCS plot corresponds to the initialization of the dynamics with the
non-localized coherent state which is a Gaussian distribution spreading over several phase eigenstates. 
A localized initial state, i.e., a single phase eigenstate, generates an interference pattern, in the
the case of optimized SCS dynamics. The localized initial states are not considered in this paper.

\begin{figure}
\includegraphics[scale=0.33]{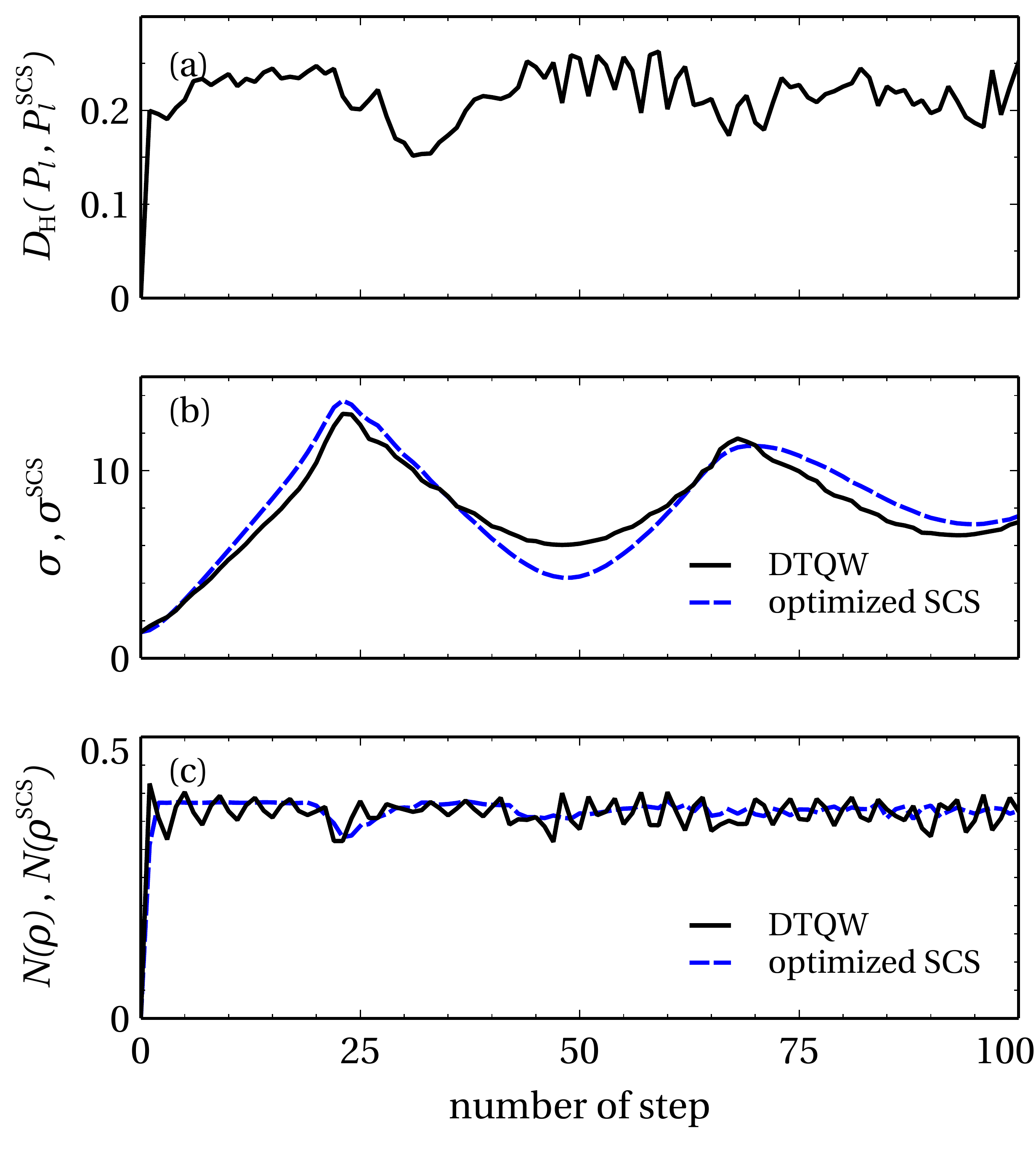}
\caption{The Hellinger distance between the phase probability distributions for the DTQW dynamics with the
Hadamard-like coin and the SCS dynamics with the optimized angles in (a), and the corresponding standard
deviations and the coin-walker entanglements in (b) and (c), respectively.
The initial coherent state is taken as $|\alpha\!=\!5e^{i\pi}\rangle$ and the phase space dimension
is set to $d\!=\! 31$. The initial state of the qubit is set to the ground state.}
\label{fig:various}
\end{figure}

After crossing the boundaries, the two peaks of either probability distributions meet and interfere.
Figure \ref{fig:various} (a) shows the Hellinger distance between the DTQW and optimized SCS probability
distributions for $100$ steps. Although optimized for the first $50$ steps, the Hellinger distance
does not increase in the next $50$ steps. Note that the Hellinger
distance between the plots in Fig.~\ref{fig:prob_dist} is about $23$\%, and the maximum distance in
Fig. \ref{fig:various} (a) is about $26$\%.

It should be mentioned that the phase probability distribution for the SCS operator with non-optimized
parameters (not shown here) consists of a single peak which is just translated in each step of the
dynamics. As we described earlier in this section, such situation was expected, since the effect of
the non-optimized SCS operator is approximately similar to the effect of the shift operator.

Having tuned the SCS operator such that it generates a phase probability distribution similar to the
DTQW dynamics, we compare, in the following, the dynamics of the corresponding standard deviations and
the coin-walker entanglements.

The standard deviation of the DTQW dynamics increases linearly in terms of time steps. We observe
the same linear growth for the optimized SCS dynamics.
Figure \ref{fig:various} (b) shows the standard deviations related to the DTQW and the optimized SCS
probability distributions.
The oscillatory behavior of the standard deviations correspond to the walk on the circle---an array with
periodic boundary conditions.

To quantify the entanglement between the coin and the walker, we use the negativity
measure \cite{vidal2002computable}
\begin{equation}
{N}(\varrho_l) = \frac{ \| \varrho_l^{T_W} \|_1 - 1 }{2},
\end{equation}
where $\varrho_l=|\psi_l\rangle\langle\psi_l|$ is the density operator of the quantum walk at step $l$,
$\varrho_l^{T_W}$ is the partial transpose of the density operator with respect to the walker and
{\color{black}
$\|.\|_1$ is the trace norm.
}
For the optimized SCS dynamics, the negativity can be obtained
by using the density operator
$\varrho^\mathrm{SCS}_l=|\psi^{\mathrm{SCS}}_l\rangle\langle\psi^\mathrm{SCS}_l|$.
Figure \ref{fig:various} (c) shows the negativity for the DTQW and the optimized SCS dynamics in $100$ steps.
It can be seen that the entanglement generated for the optimized SCS dynamics is smoother, being approximately
equal to the envelope of the entanglement plot for the DTQW dynamics.

Figure \ref{fig:various} (b) and (c), therefore, show that the optimized SCS dynamics has two important features
of the DTQW; the ballistic spread of the walker probability distribution which is manifested in the
linear behavior of the corresponding standard deviation, and the similar behavior
of the coin-walker entanglement.

\section{Decoherece in the SCS dynamics}
\label{sec:V}
The DTQW dynamics, under the effects of decoherence on the quantum coin, shows a quantum-to-classical
transition in which the probability distribution spreads diffusively, hence, the corresponding standard
deviation evolves with the square root of time steps.
We simulate the effect of the dephasing channel~\cite{nielsen2010quantum} on the coin for
the optimized SCS operator and observe similar behavior.

\begin{figure}
\includegraphics[scale=.33]{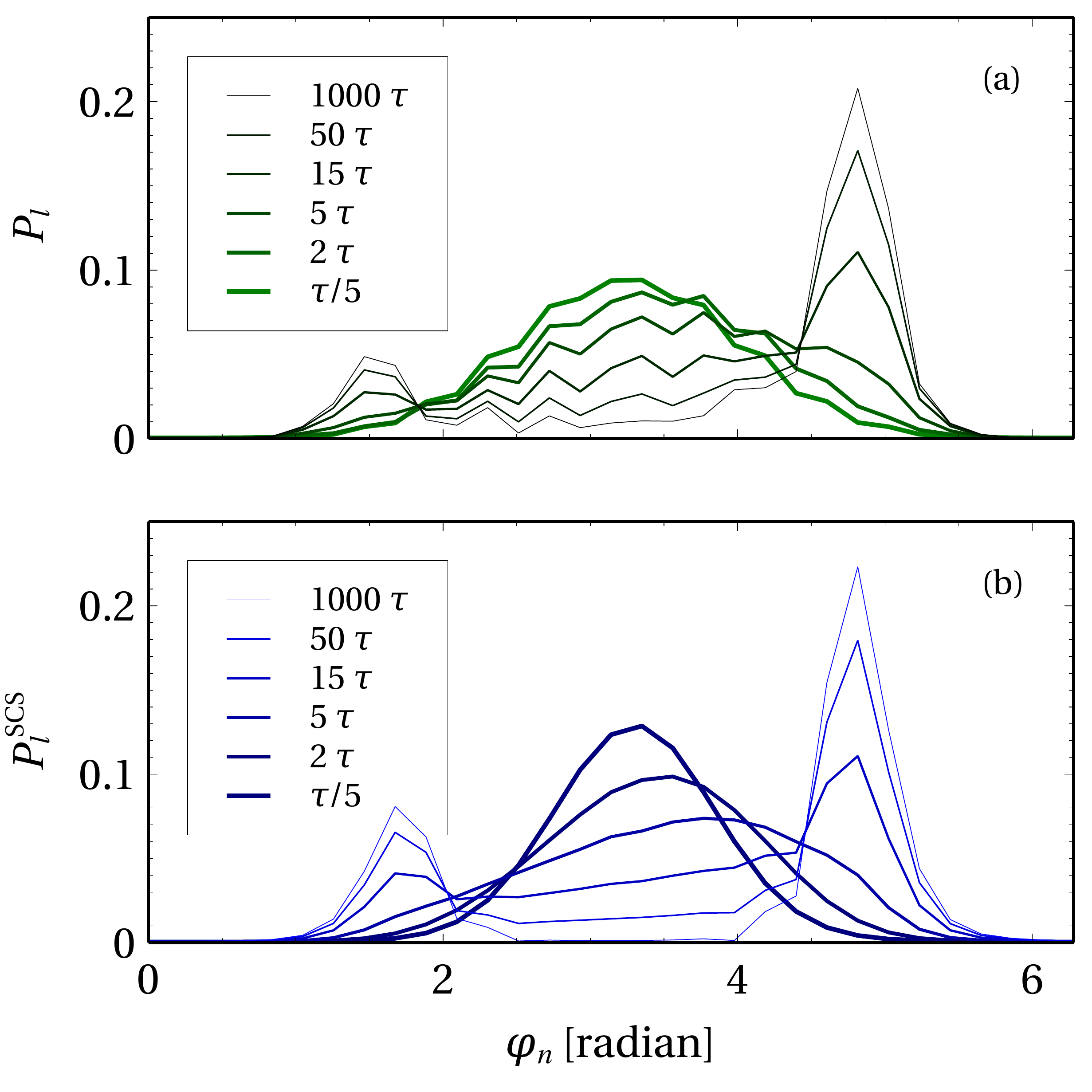}
\caption{The walker phase probability distributions at step $l=12$, for different dephasing time $T_d$
specified in terms of the time step $\tau$, for the DTQW dynamics in (a) and the optimized SCS dynamics
in (b). The dynamics is generated under the effect of the dephasing channel modeled in
Eq. (\ref{eq:phase_damp}). The initial state for either dynamics is taken as
$|\alpha\!=\!5e^{i\pi}\rangle|0\rangle$ and the phase space dimension is set to $d\!=\! 31$.}
\label{fig:decoherence}
\end{figure}

Under the effect of decoherence, the walk dynamics is not unitary and can be described
by \cite{kendon2007decoherence}
\begin{equation}
\label{eq:nonunitary_evolution}
 \varrho_l = \sum_{j} K_j U \varrho_{l-1} {{U}}^{\dagger}  K_j^{\dagger},
\end{equation}
where $\varrho_l$ is the quantum walk density operator, $K_j$ are the Kraus operators
modeling the quantum noise and ${U}$ is the DTQW operator.
Similarly, the effect of the channel on the optimized SCS dynamics is obtained by replacing $\varrho_l$ and
${U}$ in Eq. (\ref{eq:nonunitary_evolution}) with $\varrho^\mathrm{SCS}_l$ and
${U}^\mathrm{SCS}$, respectively.
The effect of the phase damping channel on the coin can be modeled by the Kraus
operators $K_j= {\mathbb{1}}\otimes E_j$ $(j=0,1$), in which ${\mathbb{1}}$ is the walker space
identity and \cite{nielsen2010quantum}
\begin{equation}
\label{eq:phase_damp}
 E_0 = \begin{pmatrix}
        1& \;\; 0 \\
        0& \;\; \sqrt{1-\lambda} \\
       \end{pmatrix},\;\;\;\;\;
 E_1 = \begin{pmatrix}
        0& \;\; 0 \\
        0& \;\; \sqrt{\lambda} \\
        \end{pmatrix},
\end{equation}
where $\lambda$ quantifies the strength of the channel and can be written as $\lambda=1 - e^{-l\tau/T_d}$,
in terms of the dephasing time $T_d$.
The system dynamics can be obtained by solving Eq.~(\ref{eq:nonunitary_evolution}) using the
initial state~(\ref{eq:initial_state}).

Figure~\ref{fig:decoherence} (a) shows the phase probability distribution for the DTQW, at step $l\!=\!12$,
for different values of the dephasing time $T_d$.
At that time step, in the case $T_d\!\to\!\infty$, the walker probability distribution forms two peaks which
are separated by an angular distance of about $\pi$ radian.
However, by decreasing $T_d$ the walker probability distribution gradually changes to a Gaussian
distribution. The corresponding dynamics of the standard deviation (not shown here) also gradually changes
from linear to square root dependency on time steps. 
Figure~\ref{fig:decoherence} (b) shows the case for the optimized SCS evolution subjected to the coin
dephasing channel. Similar quantum-to-classical transition can be observed for the optimized SCS dynamics,
when dephasing time decreases.

\begin{figure}
\includegraphics[scale=.33]{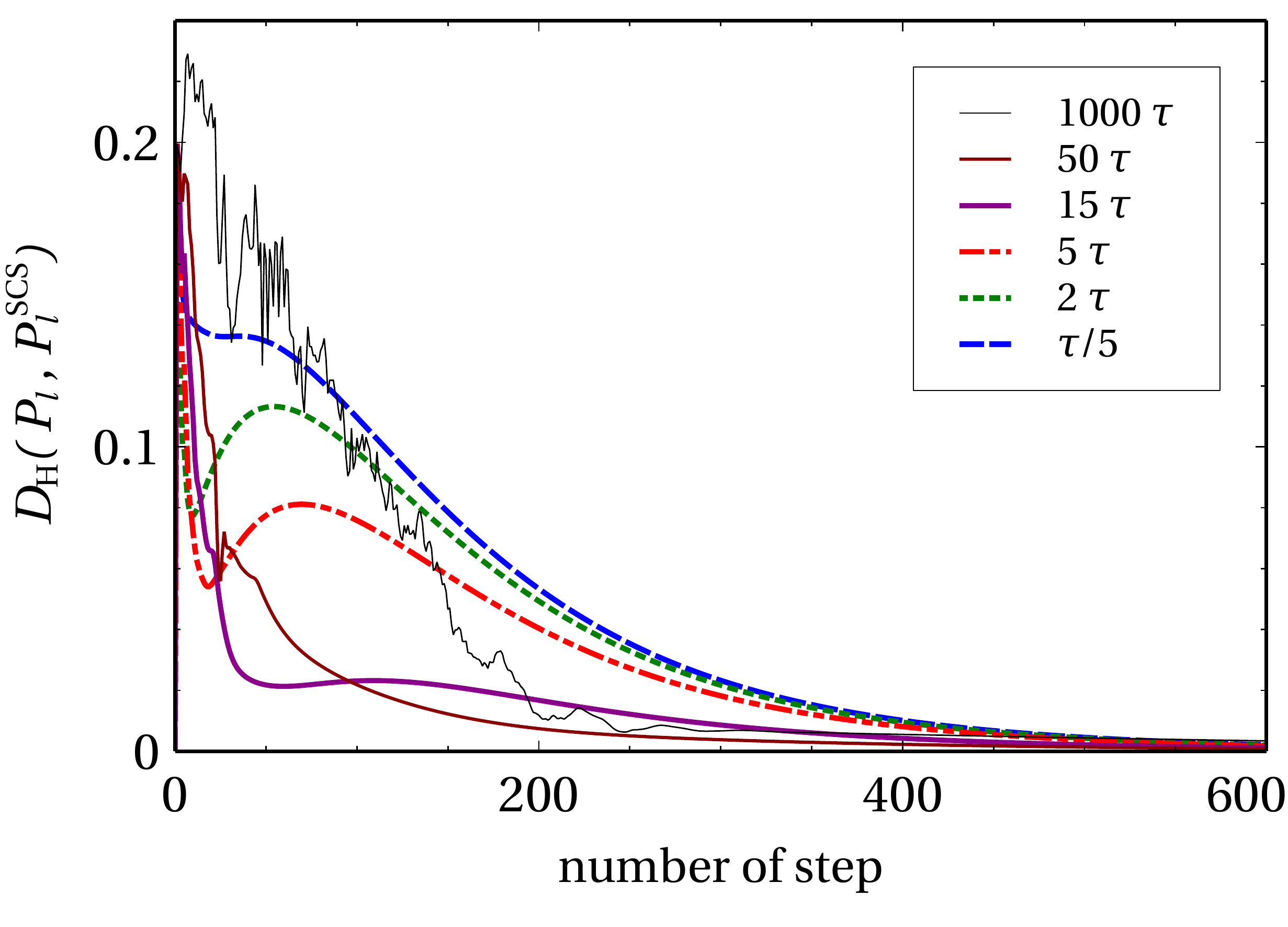}
\caption{The Hellinger distance between the phase probability distributions for the DTQW dynamics with
the Hadamard-like coin and the SCS dynamics with the optimized angles, for $600$ time steps, when either
dynamics is affected by decoherence. The dephasing times are given in terms of the time  step $\tau$.
The initial state for either dynamics is taken as
$|\alpha\!=\!5e^{i\pi}\rangle|0\rangle$ and the phase space dimension is set to $d\!=\! 31$. }
\label{fig:Hellinger_decoherence}
\end{figure}

Figure~\ref{fig:Hellinger_decoherence} shows the Hellinger distance between the phase probability
distributions related to the DTQW and the optimized SCS dynamics, for $600$ time steps, and for different
dephasing times.
It can be seen that after sufficiently large time steps the Hellinger distance approaches $0$,
independent of the dephasing times.
In fact, after sufficiently large time steps, the DTQW dynamics, for any finite dephasing time $T_d$,
makes a transition to the classical regime. In that regime, the coin is classical, hence, the dynamics
is a random walk whose limiting probability distribution converges to the uniform distribution $1/d$.
It is implied that, then, the limiting probability distribution for to the optimized SCS dynamics also
converges to the uniform distribution $1/d$. Therefore, in the classical regime there is no difference
between the DTQW and the optimized SCS operators, reflecting the commutativity of the classical operators.

Figure~\ref{fig:Hellinger_decoherence} also shows an increase in the Hellinger distance after some
time steps in the beginning, especially, when the dephasing time is smaller than a few tens
of $\tau$.
In this case, after some initial time steps, both DTQW and optimized SCS operators are
transformed to classical random walks, but, the resulting random walks are not quite similar.
That corresponds to the difference between the parameters $(\omega\tau,g\tau)$, relating to the
DTQW and the optimized SCS operators. That situation can be more clear by comparing
Figs. \ref{fig:decoherence} (a) and (b), in which after some time steps the probability distributions,
for small dephasing times, transform to Gaussian distributions but with different spreading speeds.

\section{Final considerations}
\label{sec:VI}
We analyzed the SCS dynamics which is generated by the time-independent Hamiltonian including both the coin
and the shift generators in all the times, to simulate the Hadamard DTQW dynamics.
The DTQW is generated by the time-dependent Hamiltonian alternating between the coin and the shift terms.
Both dynamics comprise the momentum-dependent rotations of the walker spin. However, their
spectrum is different and the corresponding Bloch vectors have different behavior in the Brillouin zone.
The winding number for the DTQW effective Hamiltonian is $1$ but for the SCS Hamiltonian is $0$.
Moreover, the main effect of the (non-optimized) SCS operator is just to translate the initial quantum
walk state on the array sites.

By minimizing the Hellinger distance between the phase probability distributions, related to the DTQW
and SCS dynamics, over the system frequencies, we made the SCS dynamics behave approximately similar to
the DTQW. The optimized SCS operator has a more flat spectrum and the corresponding Bloch vectors are
pushed toward the points $(\pm1,0,0)$, comparing with the non-optimized case. Moreover, the optimized
SCS operator transforms a single basis vector to a superposition of the basis vectors.

We compared the dynamics of the standard deviation and the entanglement regarding the DTQW and SCS dynamics
and found similarities between them. By using a decoherent coin, we observed that the
quantum-to-classical transition also happens for the SCS dynamics, when the decoherence strength increases.
Moreover, in the classical regime, both DTQW and SCS dynamics show random walk behavior.

The implementation of the SCS dynamics is less challenging than the DTQW, since the SCS Hamiltonian is
time-independent. The SCS Hamiltonian can be realized in a circuit QED composing of a transmission
line resonator coupled to a transmon qubit.
The system in the large detuning regime, derived by a microwave field and in the frame rotating with
the drive frequency, can be described by the Hamiltonian \cite{blais2004cavity}
\begin{align}
 \label{eq:CQED}
 H = &\frac{1}{2} \left[\omega_\mathrm{q}-\omega_\mathrm{d}+
     2\frac{g_\mathrm{qr}^2}{\Delta}(a^{\dagger}a+1/2)\right]\sigma_z+
      \frac{g_\mathrm{qr}\varepsilon(t)}{\Delta}\sigma_{x}\nonumber \\
     &+(\omega_\mathrm{r}-\omega_\mathrm{d})a^{\dagger}a+ 
      \varepsilon(t)(a^{\dagger}+a),
\end{align}
where $\omega_\mathrm{q}$ is the qubit transition frequency, $\omega_\mathrm{r}$ is the resonator frequency,
$g_\mathrm{qr}$ is the qubit-resonator coupling, $\omega_d$ is the deriving frequency,
$\Delta\!=\!\omega_\mathrm{q}-\omega_\mathrm{r}$ is
detuning and $\varepsilon(t)$ is the drive amplitude
{\color{black}
(see Appendix \ref{app:circuit_QED} for more details).
}
By setting $\omega_d\!=\!\omega_q+g_\mathrm{qr}^2/\Delta$
the required terms for the SCS Hamiltonian, namely $(g_\mathrm{qr}^2/\Delta)a^{\dagger}a\sigma_z$ and
$(g_\mathrm{qr}\varepsilon/\Delta)\sigma_{x}$ are obtained.
The term $(\omega_\mathrm{r}-\omega_\mathrm{d})a^{\dagger}a$
is just introducing a free rotation to the coherent state of the resonator, and the unwanted effects of
the term $\varepsilon(a^{\dagger}+a)$ can be reduced by decreasing $\varepsilon$.

In this paper, we have analyzed the simulation of the DTQW only with an unbiased coin, namely the Hadamard walk
which is crucial for quantum search algorithms~\cite{Portugal2013quantum} and interesting for quantum
simulations~\cite{obuse2015unveiling}.
{\color{black}
A similar approach, however, can be applied to simulate a DTQW with an arbitrary coin operator whose angle of rotation is in the interval $\theta\in[0,\pi/2]$. The optimization problem (\ref{eq:optimization}) has already been solved by considering the Hellinger distance between the SCS dynamics and the DTQW evolution with the Hadamard coin $C(\pi/4)$. Now, a new optimization problem can be solved by modifying the Hellinger distance to correspond to the coin operator $C(\theta)$ given by Eq. (\ref{eq:coin}).
Figure \ref{fig:fid_theta} shows the average Hellinger distance $\overline{D_\mathrm{H}}(P_l,P_l^\mathrm{SCS})$, obtained by averaging the Hellinger distance (\ref{eq:Hellinger}) between the phase probability distributions corresponding to the DTQW dynamics with the coin operator $C(\theta)$ and the related SCS dynamics with the optimized angles, for different values of $\theta$.
As can be seen in the figure, for $\theta \to 0$ the average distances between the two dynamics decrease which is consistent with the Trotter's relation (\ref{eq:trotter}). However, for large values of $\theta$, the two dynamics become different, hence, the difference between them cannot be decreased arbitrarily, rather it remains bounded in the whole interval $[0,\pi/2]$.

\begin{figure}
\begin{center}
\includegraphics[scale=.33]{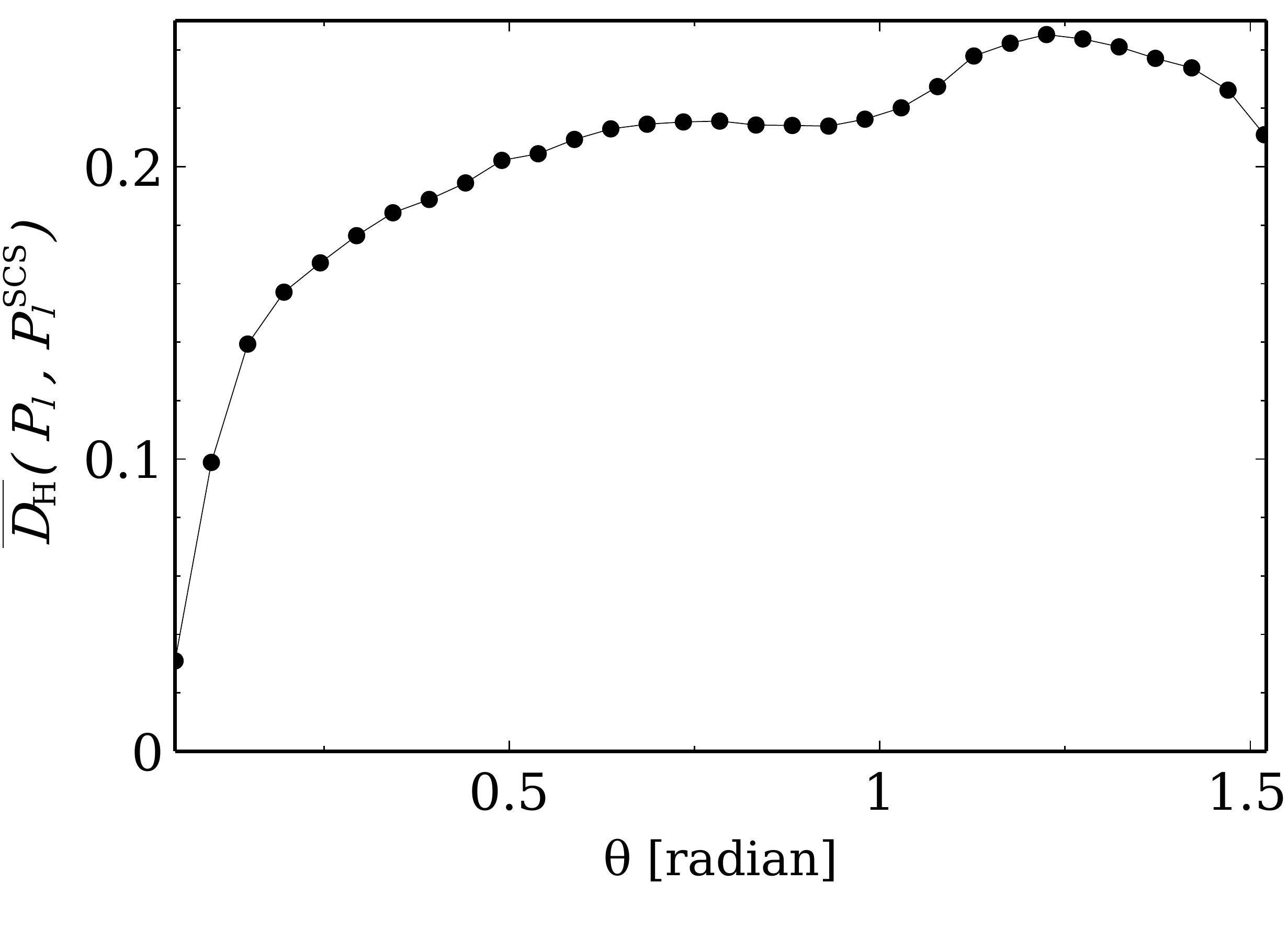}
\caption{The average Hellinger distance between the phase probability distributions for the DTQW dynamics with
the coin operator $C(\theta)$ and the SCS dynamics with the corresponding optimized angles, for $\theta$ in the interval $[\pi/64,31\pi/64]$ at $\pi/64$ steps. The initial state for either dynamics is taken as
$|\alpha\!=\!5e^{i\pi}\rangle|0\rangle$ and the phase space dimension is set to $d\!=\! 31$. }
\label{fig:fid_theta}
\end{center}
\end{figure}
}

{\color{black}
As can be inferred from Fig. \ref{fig:diag}, the dominant elements of the SCS operator are confined to a diagonal band implying that the SCS operator can be approximated by a local operator. Considering moreover the unitarity of the operator, an extension of the present work is to derive the SCS dynamics starting from the quantum cellular automata model.
}
The applications of the SCS dynamics in quantum information processing and quantum simulation needs further investigation, which should be addressed somewhere else.

\textit{Acknowledgments.}---
JKM acknowledges financial support from Iran's National Elites Foundation, grant No. 7000/2000-1396/03/08.
MCO acknowledges supports by the Funda\c{c}\~ao de Amparo \`a Pesquisa do Estado de S\~ao Paulo (FAPESP)
through the Research Center in Optics and Photonics (CePOF).

\appendix

\renewcommand{\theequation}{A\arabic{equation}}
\setcounter{equation}{0}
\renewcommand{\thefigure}{A\arabic{figure}}
\setcounter{figure}{0}
\section{Shift operator in the Fourier basis}
\label{app:app1}
The action of the Shift operator (\ref{eq:shift}) on the Fourier basis is calculated as
\begin{align}
S\;|\tilde{k}\rangle \otimes |s\rangle
&= S \frac{1}{\sqrt{d}}\sum_n e^{- i \tilde{k} n } |n\rangle \otimes |s\rangle \nonumber\\
&= \frac{1}{\sqrt{d}}\sum_n e^{- i \tilde{k} n } S|n\rangle \otimes |s\rangle \nonumber\\
&= \frac{1}{\sqrt{d}}\sum_n e^{- i \tilde{k} n } |n+(-1)^s\rangle \otimes |s\rangle \nonumber\\
&= \frac{1}{\sqrt{d}}\sum_{n'} e^{- i \tilde{k} [n' - (-1)^s] } |n'\rangle \otimes |s\rangle \label{eq:shift_action1}\\
&= e^{i \tilde{k} (-1)^s } \frac{1}{\sqrt{d}}\sum_{n'} e^{- i \tilde{k} n' } |n'\rangle \otimes |s\rangle \nonumber\\
&= e^{i \tilde{k} (-1)^s } |\tilde{k}\rangle \otimes |s\rangle, \label{eq:shift_action2}
\end{align}
where in (\ref{eq:shift_action1}) $n$ is replaced by $n' = n + (-1)^s$, and Eq. (\ref{eq:shift_action}) in Sec. \ref{sec:II} is then proved. The shift operator $S$ is diagonal in the Fourier basis, hence,
\begin{align}
\label{eq:shift_operator1}
S &= \sum_k \biggl(|\tilde{k}\rangle \langle\tilde{k}| \otimes |0\rangle \langle 0| e^{i\tilde{k}} +
   |\tilde{k}\rangle \langle\tilde{k}| \otimes |1\rangle \langle 1| e^{-i\tilde{k}} \biggr) \nonumber\\
  &= \sum_k |\tilde{k}\rangle \langle\tilde{k}| \otimes \biggl( e^{i\tilde{k}} |0\rangle \langle 0| +
                                                                e^{-i\tilde{k}} |1\rangle \langle 1| \biggr) \nonumber\\
  &= \sum_k |\tilde{k}\rangle \langle\tilde{k}| \otimes
  \begin{pmatrix}
  e^{i\tilde{k}} & 0 \\ 0 & e^{-i\tilde{k}}
  \end{pmatrix}   \nonumber\\
  &= \sum_k |\tilde{k}\rangle \langle\tilde{k}| \otimes e^{i\tilde{k}\sigma_z},
\end{align}
proving Eq. (\ref{eq:shift_operator}) in Sec. \ref{sec:II}.

\renewcommand{\theequation}{B\arabic{equation}}
\setcounter{equation}{0}
\renewcommand{\thefigure}{B\arabic{figure}}
\setcounter{figure}{0}
\section{Bloch vectors}
\label{app:bloch_vectors}
The Bloch vectors in Eq. (\ref{Bloch_vectors}) are obtained by substituting Eq. (\ref{eq:coin}) together
with
\begin{equation}
e^{i\tilde{k}\sigma_z} = \begin{pmatrix}
  e^{i\tilde{k}} & 0 \\ 0 & e^{-i\tilde{k}}
  \end{pmatrix},
\end{equation}
and
\begin{equation}
e^{-i \epsilon_\theta(\tilde{k}) \bm{d}_\theta(\tilde{k}).\bm{\sigma}} =
\cos \epsilon_\theta(\tilde{k}) \; \mathbb{1}_2 - i \sin \epsilon_\theta(\tilde{k})
\begin{pmatrix}
  d_z & d_x-id_y \\ d_x+id_y & -d_z \\
  \end{pmatrix},
\end{equation}
in Eq. (\ref{eq:general_rotation})
where $\bm{d}_\theta(\tilde{k})$ is supposed to be a unite vector with the components $(d_x,d_y,d_z)$ and $\mathbb{1}_2$ is the identity of the two-dimensional spin Hilbert space. Equation (\ref{eq:general_rotation}) also gives the dispersion relation.

\renewcommand{\theequation}{C\arabic{equation}}
\setcounter{equation}{0}
\renewcommand{\thefigure}{C\arabic{figure}}
\setcounter{figure}{0}
\section{Circuit QED}
\label{app:circuit_QED}
The Hamiltonian of the system is given by the Jaynes-cummings model including the terms corresponding to the two-level atom, the quantized field and the atom-field interaction \cite{blais2004cavity}
\begin{equation}
 \label{eq:hamiltonian_Jayness_Cummings}
 {H}_\mathrm{JC} = \frac{1}{2} \omega_q \sigma_z
                   + \omega_r a^{\dagger} a
                   + g_\mathrm{qr} (a^{\dagger} \sigma^- + a \sigma^+),
\end{equation}
where $\sigma^+$ ($\sigma^-$) is the rising (lowering) operator of the atom. To control the state of the qubit (realizing the spin rotation), the system is irradiated by a microwave field with a frequency close to the qubit's frequency
\begin{equation}
 \label{eq:hamiltonian_drive}
 {H}_d = \varepsilon(t)( a^{\dagger}e^{-i\omega_dt} + ae^{i\omega_dt} ).
\end{equation}
In the large detuning regime ($g_\mathrm{qr}\!\ll\!\Delta\!=\!\omega_q\!-\!\omega_r$), by applying the unitary transformation
$ U = e^{(a\sigma^+ - a^{\dagger}\sigma^-)g_\mathrm{qr}/\Delta}$
on the total Hamiltonian and expanding the result up to the second order in $g/\Delta$, we obtain
\begin{align}
 U ({H}_\mathrm{JC} + {H}_d) U^{\dagger} \approx&\;
 \frac{1}{2} (\omega_q + \frac{g_\mathrm{qr}^2}{\Delta}) \sigma_z +
 (\omega_r + \frac{g_\mathrm{qr}^2}{\Delta} \sigma_z) a^{\dagger} a + \nonumber \\
 &\varepsilon(t) \left[ (a^{\dagger} + \frac{g_\mathrm{qr}}{\Delta} \sigma^{+}) e^{- i \omega_d t} +
 (a + \frac{g_\mathrm{qr}}{\Delta} \sigma^{-}) e^{  i \omega_d t}\right].
\end{align}
Switching to the frame rotating at the drive frequency $\omega_d$ by applying the transformation $V=e^{i \omega_d a^{\dagger}at}$ leads to Eq. (\ref{eq:CQED}) in Sec \ref{sec:VI}.

\bibliographystyle{qinp}
\bibliography{zbib}

\end{document}